\documentclass[amsmath,amssymb,floatfix,lengthcheck,nopreprintnumbers,prl,showpacs,superscriptaddress,twocolumn]{revtex4}
\usepackage{graphicx}

\newcommand{\beq}{\begin{equation}}
\newcommand{\eeq}{\end{equation}}
\newcommand{\beqs}{\begin{eqnarray}}
\newcommand{\eeqs}{\end{eqnarray}}

\newcommand{\pbp}[0]{\ensuremath{\langle \overline{\psi} \psi \rangle}}
\newcommand{\qbq}[0]{\ensuremath{\langle \overline{q} q \rangle}}

\newcommand{\MSbar}[0]{\overline{\text{MS}}}

\begin{document}

\title{Toward TeV Conformality}

\author{T.~Appelquist}
\affiliation{Department of Physics, Sloane Laboratory, Yale University,
             New Haven, Connecticut 06520, USA}
\author{A.~Avakian}
\affiliation{Department of Physics, Boston University,
	Boston, Massachusetts 02215, USA}
\author{R.~Babich}
\affiliation{Department of Physics, Boston University,
	Boston, Massachusetts 02215, USA}
\author{R.~C.~Brower}
\affiliation{Department of Physics, Boston University,
	Boston, Massachusetts 02215, USA}
\author{M.~Cheng}
\affiliation{Physical Sciences Directorate, Lawrence Livermore National Laboratory,
	Livermore, California 94550, USA}
\author{M.~A.~Clark}
\affiliation{Harvard-Smithsonian Center for Astrophysics, Cambridge, Massachusetts 02138, USA}
\affiliation{Initiative in Innovative Computing, Harvard University School of Engineering and Applied Sciences, Cambridge, Massachusetts 02138, USA}
\author{S.~D.~Cohen}
\affiliation{Department of Physics, Boston University,
	Boston, Massachusetts 02215, USA}
\author{G.~T.~Fleming}
\affiliation{Department of Physics, Sloane Laboratory, Yale University,
             New Haven, Connecticut 06520, USA}
\author{J.~Kiskis}
\affiliation{Department of Physics, University of California,
	Davis, California 95616, USA}
\author{E.~T.~Neil}
\affiliation{Department of Physics, Sloane Laboratory, Yale University,
             New Haven, Connecticut 06520, USA}
\author{J.~C.~Osborn}
\affiliation{Argonne Leadership Computing Facility,
	Argonne, Illinois 60439, USA}
\author{C.~Rebbi}
\affiliation{Department of Physics, Boston University,
	Boston, Massachusetts 02215, USA}
\author{D.~Schaich}
\affiliation{Department of Physics, Boston University,
	Boston, Massachusetts 02215, USA}
\author{P.~Vranas}
\affiliation{Physical Sciences Directorate, Lawrence Livermore National Laboratory,
	Livermore, California 94550, USA}
\collaboration{Lattice Strong Dynamics (LSD) Collaboration}
\noaffiliation

\begin{abstract}

 We study the chiral properties of
 an $SU(3)$ gauge theory with $N_f$ massless Dirac fermions in the fundamental representation when $N_f$ is increased from $2$ to $6$. For $N_f=2$, our lattice simulations lead to a value of $\langle \bar{\psi} \psi \rangle/F^3$, where $F$ is the Nambu-Goldstone-boson decay constant and $\langle \bar{\psi} \psi \rangle$ is the chiral condensate, which agrees with the measured QCD value. For $N_f = 6$, this ratio shows significant enhancement, presaging an even larger enhancement anticipated as $N_f$ increases further, toward the critical value for transition from confinement to infrared conformality.
\end{abstract}


\maketitle

\paragraph{\textbf{Introduction}}

Theories with an approximate conformal symmetry could play a role in describing
new physics at the TeV scale and beyond. While
a nonsupersymmetric, vectorlike gauge theory exhibits
confinement and spontaneous chiral symmetry breaking with a small number $N_f$ of
massless fermions, it becomes conformal in the infrared, governed by a weak infrared
fixed point if $N_f$ is larger, but just below the value for which asymptotic
freedom sets in \cite{Caswell:1974gg}. There is evidence from lattice simulations \cite{Appelquist:2007hu, Deuzeman:2008sc, Appelquist:2009ty, Deuzeman:2009mh, Fodor:2009wk, Nagai:2009ip} that this infrared conformality persists down through a ``conformal window" of $N_f$ values
where the fixed point can become strong, and that a transition to the confining and chirally broken phase takes place at some value $N_f^c$.

Even for $N_f < N_f^c$ there can remain an approximate infrared fixed point providing that $0 < N_f^c - N_f \ll N_f^c$. The scale $F$ of chiral symmetry breaking is then small relative to the intrinsic scales of the theory, and the fixed point approximately governs the theory from the breaking scale out to some higher scale.

This ``walking" phenomenon can play an important phenomenological role in a technicolor theory of electroweak symmetry breaking.
Flavor-changing neutral currents (FCNC's), which are present when the technicolor theory is extended to provide for the generation of quark masses, can be too large unless the associated scale $\Lambda_{ETC}$ is high enough. But then the first- and second-generation quark masses are typically much too small. They are proportional to the quantity $\pbp /\Lambda_{ETC}^2$, where $\psi$ is a technifermion field and $\pbp$ is the bilinear fermion condensate defined (cut off) at $\Lambda_{ETC}$.
Walking can lift the quark masses by enhancing the condensate
  $\pbp$ significantly above its value
 [$O(4 \pi F^3)$] in a QCD-like theory \cite{Holdom:1981rm,Yamawaki:1985zg,Appelquist:1986an},
  while keeping $\Lambda_{ETC}^2$ large enough to suppress FCNC's.

The enhancement of $\pbp / F^3$ as $N_f \rightarrow N_f^c$ from below has been indicated by Feynman-graph-based  studies. But it is important also to use nonperturbative methods since the couplings involved are strong. This Letter describes a first step in this program. We focus on an $SU(3)$ gauge theory with $N_f$ massless Dirac fermions in the fundamental representation.  Lattice studies have shown that the $N_f = 8$ theory is chirally broken, with no evidence for even an approximate infrared fixed point \cite{Appelquist:2007hu,Appelquist:2009ty, Nagai:2009ip}; while there is lattice evidence for conformal behavior at $N_f = 12$, indicating that $8 < N_f^c < 12$ \cite{Appelquist:2007hu,Deuzeman:2008sc,Appelquist:2009ty,Deuzeman:2009mh,Fodor:2009wk}.

 We present results here for the values $N_f = 2$ and $N_f = 6$, drawing on newly available computational resources, including $150 \times 10^6$ core hours on the BlueGene/L supercomputer at
 Lawrence Livermore National Laboratory (LLNL). Starting with $N_f = 2$ allows us to check the reliability of our methods by comparison with the phenomenological value of $\pbp / F^3$
for QCD. Proceeding carefully toward $N_f^c$ is prudent since the emergence of widely separated scales associated with the approximate infrared fixed point of walking is problematic for lattice methods.

\paragraph{\textbf{Methods}}

For a range of small fermion masses $m$, we compute the Nambu-Goldstone-boson (NGB) mass $M_m$, the NGB decay constant $F_m$, and the chiral condensate per fermion $\pbp_m$. To set a physical scale, we also compute the mass $M_{\rho,m}$ of the analogue of the  $\rho$ meson and the Sommer scale $r_{0,m}$ at which $r^{2}dV(r)/dr=1.65$, where $V(r)$ is the static potential
\cite{Sommer:1993ce}. Since our goal is to search for the enhancement of $\pbp / F^3$ as $N_f \rightarrow N_f^c$, from the emergence of walking between the physical length scale and the ultraviolet cutoff, taken here to be the lattice spacing, it is important to keep the lattice spacing fixed (and small) in physical units. We first choose a value for $\beta \equiv 6/g_0^2 $ at $N_f = 6$, giving a physical scale of several lattice units. For $N_f=2$, we then tune $\beta$ to match the same physical scale in lattice units.

For small enough fermion mass (and yet large enough to insure that finite-volume effects are small), the
extrapolation $m \to 0$ can be carried out by fitting the results for $M_m^2$,  $F_m$ and $\pbp_m$ to continuum $\chi$PT. The next-to-leading-order (NLO) expressions are
\cite{Gasser:1986vb}
\beq
M^{2}_m = \frac{2m \pbp}{F^2}\left\{1 + zm \left[\alpha_{M} +  \frac{1}{N_f} \log(zm)\right]\right\},
\label{Mm}
\eeq
\beq
F_m = F \left\{1 + zm \left[\alpha_{F} - \frac{N_f}{ 2}\log(zm)\right]\right\},
\label{Fm}
\eeq
\beq
\pbp_m \!=\! \pbp \left\{\!1 \!+\! zm \!\left[\alpha_{C} \!-\! \frac{N_f^2 - 1}{N_f}\log(zm)\!\right]\!
\right\},\!
\label{Pm}
\eeq
where $z = 2\pbp/ (4\pi)^{2}F^4$. The leading terms incorporate the Gell-Mann-Oakes-Renner (GMOR) relation. These expressions will be directly useful for $N_f = 2$, but because of the growth with $N_f$ of the chiral-log terms, not for $N_f = 6$.

\paragraph{\textbf{Simulation Details}}

 We use domain-wall fermions with the Iwasaki improved gauge action, as used by the RBC-UKQCD collaboration \cite{Allton:2008pn}. Lattice fermion discretization typically breaks chiral symmetry, but in the domain-wall formulation the breaking is exponentially suppressed (with flavor symmetry preserved), making it ideal for the study of chiral dynamics.
 Gauge configurations are generated using the hybrid Monte Carlo method as implemented in the USQCD application libraries, in particular {\sc cps}, via a multilevel symplectic integrator, and using Hasenbusch preconditioning and chronological inversion. Autocorrelation is reduced by blocking over sets of 50 trajectories.

The lattice volume is set to $32^3 \times 64$, with the length of the fifth dimension $L_s = 16$ and the domain-wall height $m_0 = 1.8$. All quantities are given in lattice units. For $N_f = 6$ we choose $\beta= 2.10$. For $N_f = 2$ the choice $\beta=2.70$ then leads
to nearly the same physical scale in lattice units. Simulations are performed for fermion masses $m_f = 0.005$ to $0.03$. At finite lattice spacing, even with $m_f = 0$, the chiral symmetry is not exact, with the violation captured in a residual mass $m_{res}\ll m_f$. For $N_f = 2$, $m_{res} = 2.60 \times 10^{-5}$, while for $N_f = 6$, $m_{res} = 8.23 \times 10^{-4}$. The total fermion mass $m$ is then  $ m \equiv m_f + m_{res}$.

Although global topological charge, $Q$, is an irrelevant quantity with infinite volume, in a finite volume it becomes relevant \cite{Leutwyler:1992yt}. On a discretized lattice, $Q$ is not conserved, with the system evolving between sectors, an evolution crucial for the correct sampling of the path integral at finite volume. With very light fermions, the evolution of $Q$ slows dramatically using current Monte Carlo methods \cite{Jung:2009}. We find that $Q$ evolves sufficiently for $m_f \geq 0.01$ for both $N_f=2$ and 6. At $m_f = 0.005$ it does not, leading to systematic shifts in $\pbp_m$ and $F_m$, which we will explore in a future paper. Here, we present results for $m_f = 0.005$, but do not include them in our analysis. This also ensures that for each $m$,  $M_{m}L > 4$, keeping the NGB Compton wavelength well inside the lattice (the p regime). Results for $m_f = 0.025, 0.03$ are likewise not used in our analysis.

\paragraph{\textbf{Results}}

We first report our results for the extrapolated values of the physical scales $M_{\rho,m}$ and $1/r_{0,m}$, with $\beta= 2.10$ at $N_f = 6$ and $\beta=2.70$ at $N_f = 2$.  For small enough $m$, they can be extrapolated to $m = 0$ using $\chi$PT \cite{Leinweber:2001ac}, where the NLO terms are now linear in $m$ (there is no $m$ log($m$) term). This and
the small change in these quantities in the range $m = 0.01$ -- $0.02$, indicate that a linear extrapolation should suffice for both $N_f = 2$ and $N_f = 6$. The extrapolated values in lattice units are $r_{0}^{-1} = 0.111(4)$ ($N_f = 2$) and $0.100(6)$ ($N_f = 6$), and $M_{\rho} = 0.198(14)$ ($N_f = 2$) and $0.207(15)$ ($N_f = 6$). Thus, to within the $10\%$ accuracy of this Letter, both $r_{0}^{-1}$ and $M_{\rho}$ remain fixed going from $N_f = 2$ (with $\beta=2.70$) to $N_f = 6$ (with $\beta= 2.10$). We note finally that
for QCD, $r_0 = 0.378(9)\ \textrm{GeV}^{-1}$ \cite{Gray:2005ur}, giving $1/M_{\rho} r_0 = 0.488(12)$, in reasonable agreement with our $N_f = 2$  value $1/M_{\rho} r_0 = 0.561(44)$.

\begin{figure}
\includegraphics[width=85mm]{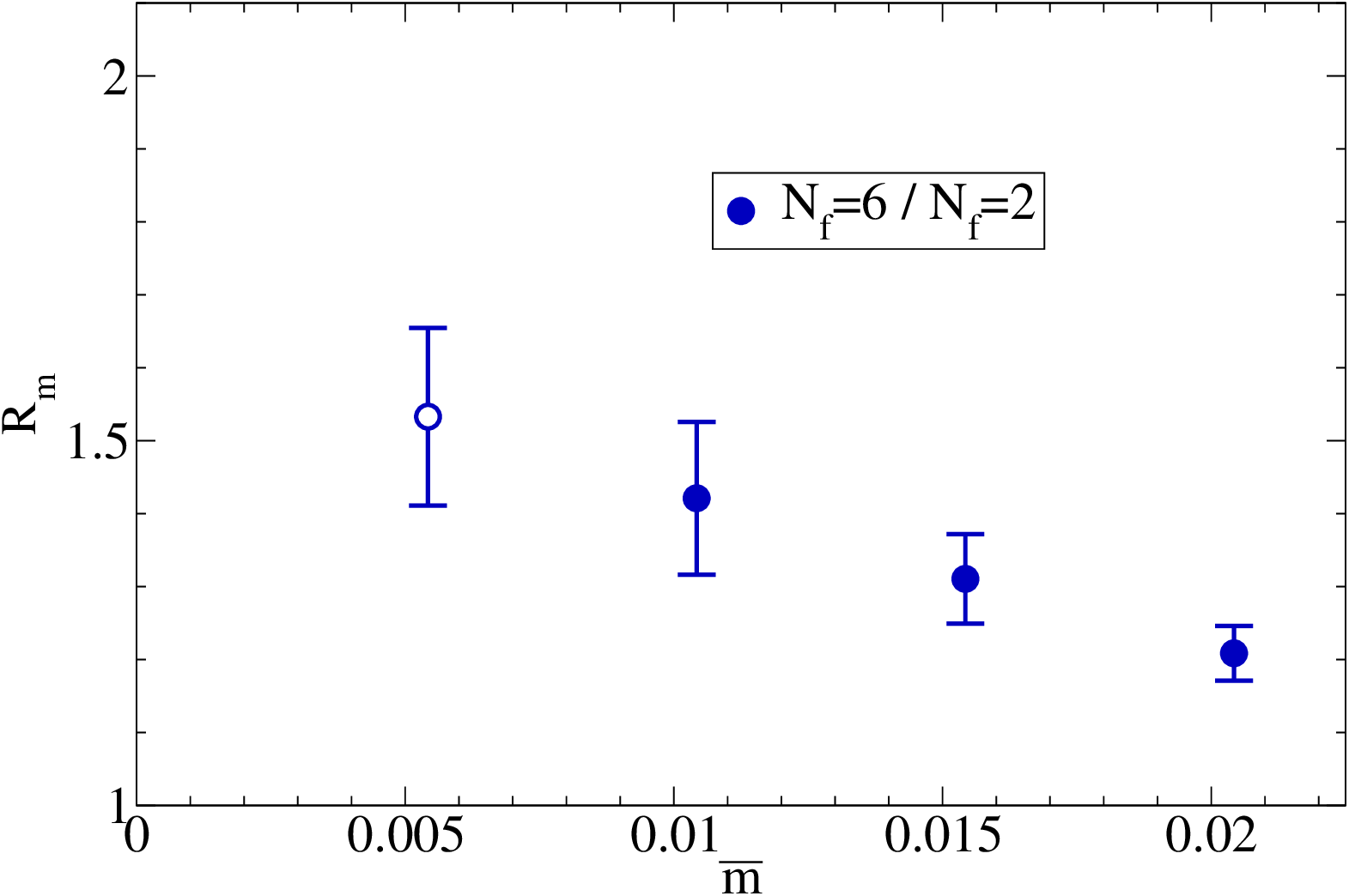}
\caption{\label{ratio} $R_m \equiv [M_m^{2}/2mF_m]_{6f}/[M_m^{2}/2mF_m]_{2f}$, versus $\overline{m} \equiv (m_{2f} + m_{6f})/2$, indicating enhancement of $\pbp/F^3$ at $N_f = 6$ relative to $N_f = 2$. \vspace{-2mm}}  
\end{figure}

We now turn to $M_m^2$, $F_m$, and $\pbp_m$, noting first that $M_m^2/2mF_m$  extrapolates to $\pbp / F^3$ (the GMOR relation) in the chiral limit. We can get an estimate of the enhancement of this ratio by
 comparing $M_m^2/2mF_m$ for $N_f = 6$ to that for $N_f = 2$, at finite $m_f$. We do this by plotting the ratio of ratios $R_m \equiv [M_m^{2}/2mF_m]_{6f}/[M_m^{2}/2mF_m]_{2f}$ in Fig.~\ref{ratio}. We use the ratio $M_m^2/2mF_m$ here, but we could also use the ratios $\pbp_m/F_m^3$ or
 $(M_m^2/2m)^{3/2}/\pbp_m^{1/2}$, which also extrapolate to $\pbp / F^3$, and which
 show the same trend. The evident trend in Fig.~\ref{ratio} is that $R_m$ increases as $m_f$ decreases. Even disregarding the point at $m_f = 0.005$, this suggests that the extrapolated value will be well above unity. A linear extrapolation of the data for $m=0.01 - 0.02$ gives
 $1.58(19)$ for $m=0$, implying an enhancement in this range or above unless there is a downturn in $R_m$. This would require either special values of the $\chi$PT  parameters given the natural upturn of the combined chiral-log terms in Eqs. \ref{Mm} and \ref{Fm}, or a significant downturn before $\chi$PT turns the curve up again as $m \to 0$.

\begin{figure}
\includegraphics[width=85mm]{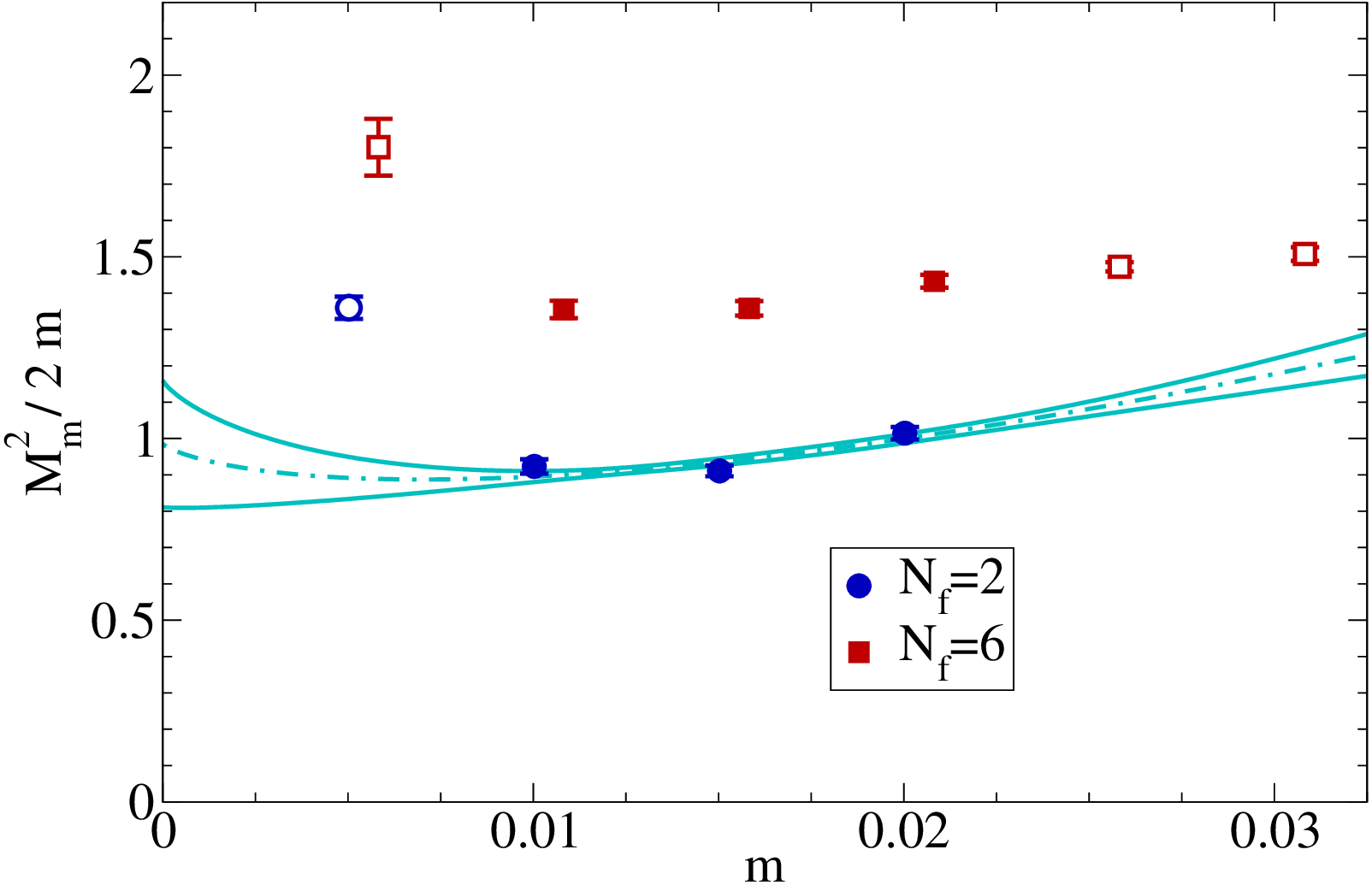}
\caption{\label{fig:B_extrap}
The slope of the pseudoscalar mass squared $M_m^2/2m$ in lattice units, as a function of fermion mass. The fit for $N_f = 2$ is a joint fit to $M_m^2$, $F_m$ and $\pbp_m$, using the (solid) points at $m_f = 0.01 - 0.02$, constrained to match NLO $\chi$PT.}

\end{figure}

\begin{figure}
\includegraphics[width=85mm]{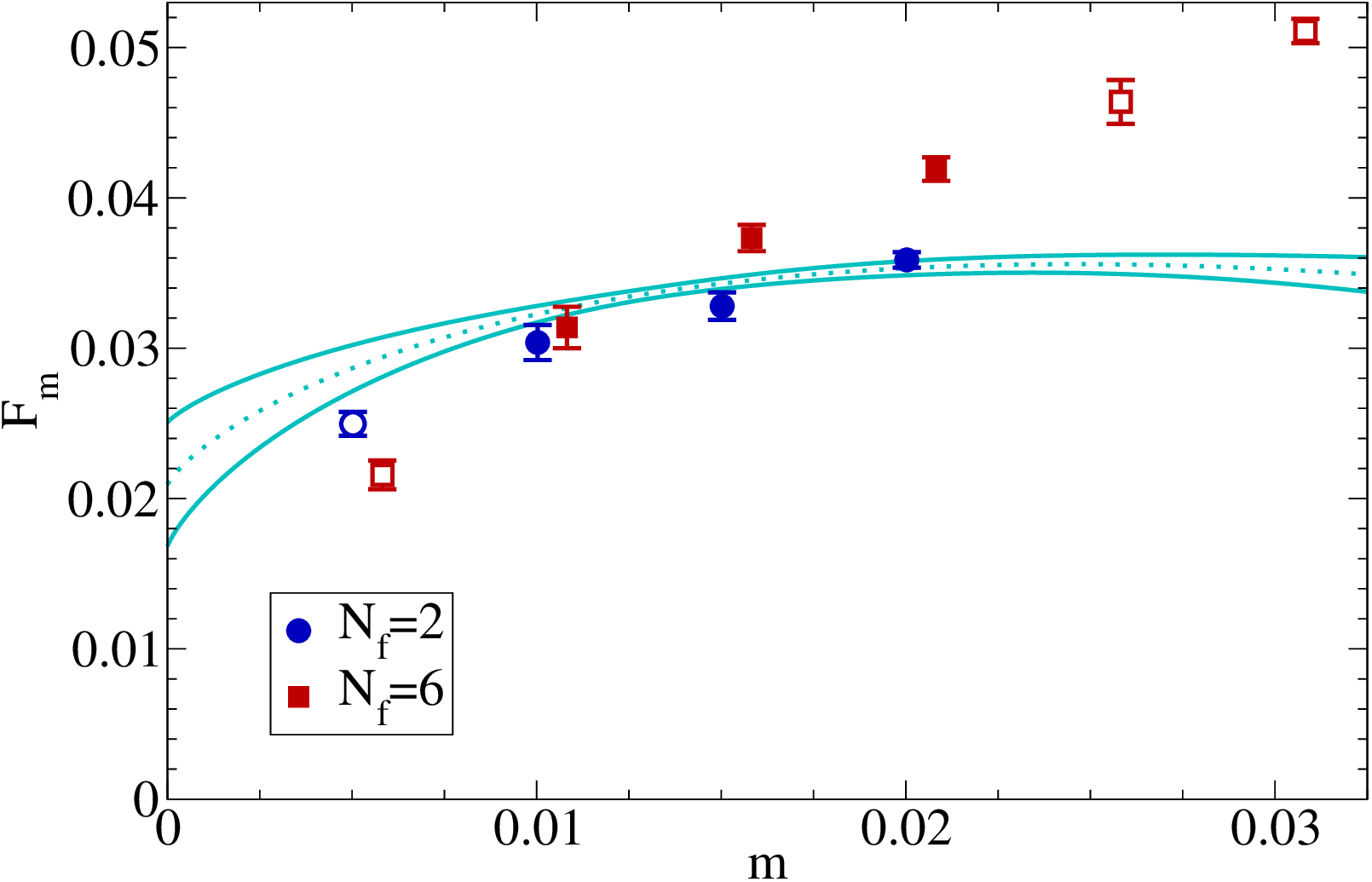}
\caption{\label{fig:F_extrap}
The Goldstone-boson decay constant $F_m$ in lattice units, as a function of fermion mass.  The fit for $N_f = 2$ is a joint fit to $M_m^2$, $F_m$ and $\pbp_m$, using the (solid) points at $m_f = 0.01 - 0.02$, constrained to match NLO $\chi$PT.}
\end{figure}

\begin{figure}
\includegraphics[width=85mm]{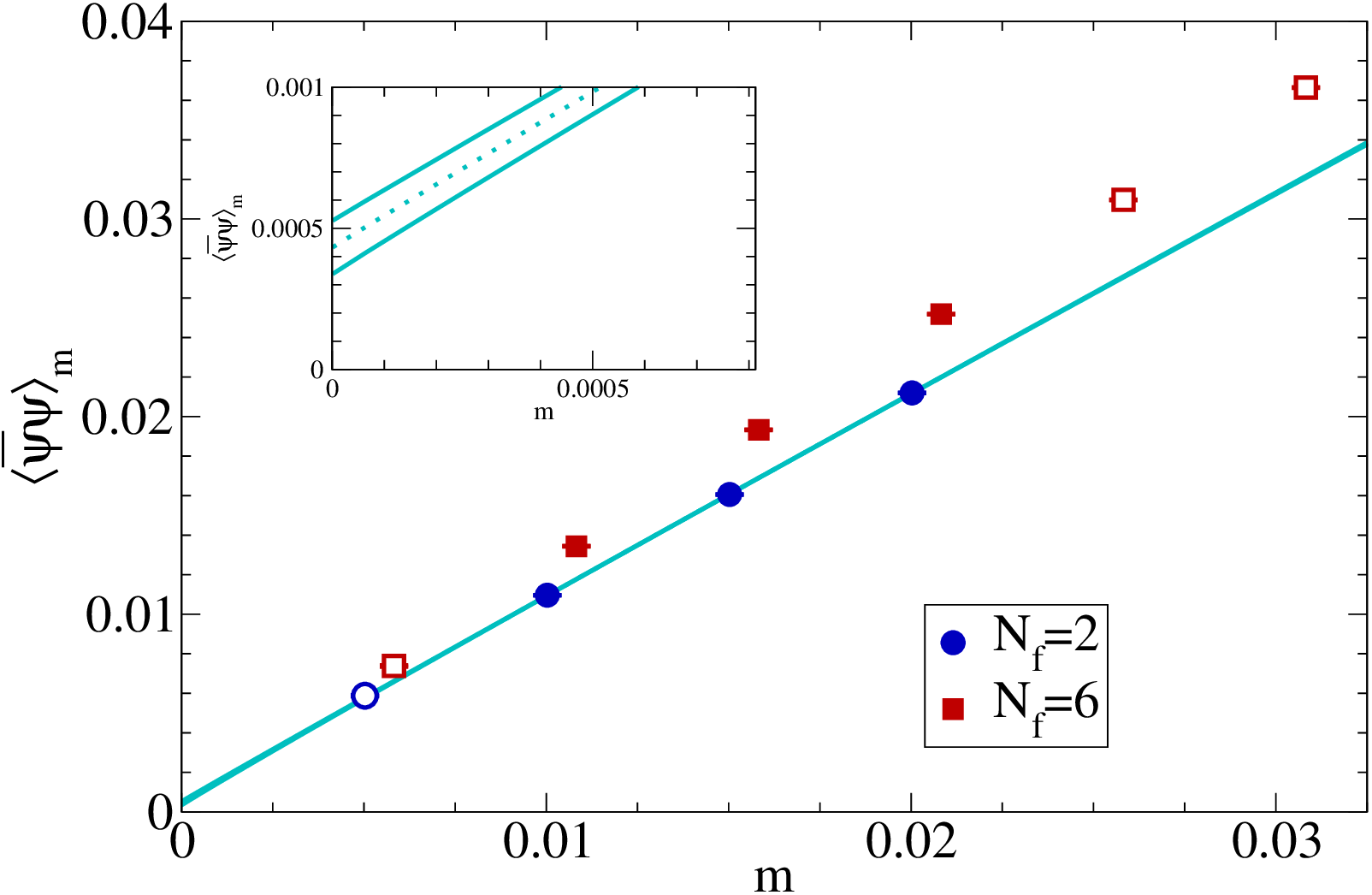}
\caption{\label{fig:pbp_extrap}
The chiral condensate per fermion $\pbp_m$ in lattice units, as a function of fermion mass. The fit for $N_f = 2$ is a joint fit to $M_m^2$, $F_m$ and $\pbp_m$, using the (solid) points at $m_f = 0.01 - 0.02$, constrained to match NLO $\chi$PT.}
\end{figure}

The separate simulation results for $M_m^2/2m$, $F_m$, and $\pbp_m$ are shown in Figs. \ref{fig:B_extrap} --  \ref{fig:pbp_extrap}. We discuss first the evidence that for $N_f = 2$, NLO $\chi$PT (Eqs. (\ref{Mm}) -- (\ref{Pm})) can  be used for the extrapolation to $m=0$. We note that for the range $m_f = 0.01 - 0.02$, for which $M_{m}L > 4$, we have $F_{m}L = O(1)$, raising another concern about the applicability of NLO $\chi$PT. Nevertheless, it has been observed that for QCD studies
with similar $F_{m}L$ values, the finite-volume corrections are no more than a few percent
\cite{Allton:2008pn}. We will examine these finite-volume corrections in a future paper \cite{LSD:2010}, relying here on the smallness of the RBC finite-volume corrections .

From Figs. \ref{fig:B_extrap} and \ref{fig:F_extrap}, we see that for $N_f = 2$, $M_m^{2}/2m$ and $F_m$ change little in the range $m_f = 0.01 - 0.02$, indicating along with the $N_f$ dependence in Eq. \ref{Pm} that NLO $\chi$PT should provide a reliable fit. Our results for $\pbp_m$ show the expected dominance of the large linear contact term (which does not preclude the use of $\chi$PT). We therefore carry out the $N_f = 2$ extrapolation to $m = 0$ using the combined NLO chiral expansions of Eqs. (\ref{Mm}) -- (\ref{Pm}). The five-parameter fit, shown in Figs.~\ref{fig:B_extrap} - \ref{fig:pbp_extrap}, leads to extrapolated values $F = 0.0209(41)$ and $\pbp/F^2 = 0.99(17)$, giving the NLO $\chi$PT result
\beq
N_f = 2: ~~~~~~~ \frac{\pbp}{F^3}  = 47.1(17.6).
\label{2fratio}
\eeq
The other fit parameters are $\alpha_M = 0.31 (62)$, $\alpha_F = 0.64 (47)$ and $\alpha_C = 83 (29)$.
The values of the fit parameters excluding $\alpha_C$ indicate that we are 
near the limit of applicability of $\chi$PT. We note also that  $\chi^2/$d.o.f. $= 6.50$ with an
uncorrelated fit and $4$ degrees of freedom. 

We next compare our $N_f =2$ results for $\pbp/F^3$ and $M_{\rho}/F$ to the QCD quantities $\qbq/f_{\pi}^3$ and $m_{\rho}/f_{\pi}$, a reasonable comparison since the light-quark masses $m_q$ are so small. With $f_{\pi} = 92.4 (0.3)$ MeV and $m_{\rho} = 775$ MeV, we have $m_{\rho}/f_{\pi} = 8.39 (0.04)$, compared to our value $M_{\rho}/F =9.4 (2.5)$. The condensate $\qbq$ is renormalization-scheme dependent, as is $m_q$. In the $\MSbar$ scheme at $2$ GeV ($\simeq2.6 m_{\rho}$),  Ref. \cite{Jamin:2002ev} finds $\qbq_{2\ \textrm{GeV}}/f_{\pi}^3 = 24.1(4.3)$.  In our case, $\pbp$ is defined by lattice regularization with $a^{-1} \simeq 5 M_{\rho}$ (equivalent to $3.85$ GeV). The increase in $\qbq$ going to this higher scale can be estimated perturbatively from the anomalous dimension of the mass operator \cite{Vermaseren:1997fq}. We find $\qbq_{3.85\ \textrm{GeV}}/f_{\pi}^3 = 29.5(5.3)$. There is also a renormalization factor $Z^{\MSbar}$, which converts the $\MSbar$ condensate to the lattice-cutoff scheme. Using Ref. \cite{Aoki:2002iq}, we find $Z^{\MSbar}(3.85~ \textrm{GeV}) = 1.227(11)$, and therefore $\qbq_{3.85 ~\textrm{GeV}, \textrm{lat}}/f_{\pi}^3 = 36.2(6.5)$, which agrees with Eq.~\ref{2fratio} within errors.

Finally, we discuss our $N_f = 6$ results, where both the simulation data and the $N_{f}$ dependence of the NLO-$\chi$PT expressions indicate that the $m$ values are not yet small enough, and the volume not yet large enough, to validate the use of NLO $\chi$PT. (We nevertheless note that a simple polynomial fit to the solid points for $M_m^2/2m$, $F_m$, and $\pbp_m$ leads to extrapolated values that satisfy the GMOR relation at $1.3 \sigma$.)
 We argue, though, that a conservative lower bound can be indirectly placed on $\pbp /F^3$ for $N_f = 6$ by bounding $F$ from above and $\pbp /F^2$ from below.

 For $F_m$ (Fig. \ref{fig:F_extrap}), the $N_f = 6$ points at $m_f = 0.01 - 0.02$ decline steeply with decreasing $m$. The eventual reliability of $\chi$PT [Eq. \ref{Fm}] at lower masses will, because of the negative curvature in the chiral-log term, bend the points down even more rapidly. An upper bound on the extrapolated value $F$ should therefore emerge from a linear fit through the three points. This gives $F \leq 0.0208 (26)$, essentially the same as the \emph{value} of $F$ in the $N_f = 2 $ case. For $M_m^2/2m$ (Fig. \ref{fig:B_extrap}), the $N_f = 6$  points in the range $m_f = 0.01 - 0.02$ are nearly flat as a function of $m$. Since $\chi$PT behavior [Eq.\ref{Mm}], with its positive curvature, sets in either in this mass range or lower, bending the points up as $m$ is decreased, a lower bound on the extrapolated value $\pbp / F^2$ should emerge from a linear fit through these three points. This gives $\pbp / F^2 \geq 1.25 (5)$.

Together, these give the conservative lower bound
\beq
N_f = 6:~~~~~~~\frac{\pbp}{F^3}  \geq 60.0~(8.0).
\label{6fratio}
\eeq
The central value of even this conservative bound shows roughly a $30\%$ enhancement relative to the $N_f=2$ central value [Eq.\ref{2fratio}]. A comparison to the more precise $N_f = 2$ ratio from QCD phenomenology [$36.2 ~(6.5)$], leads to an increase of at least $30 \%$ at $1\sigma$. And we note that these bounds are well below the enhancement indicated by inspection of Fig. \ref{ratio}.

\paragraph{\textbf{Conclusion}}

The ratio $\pbp /F^3$ in an $SU(3)$ gauge theory with $N_f$ massless Dirac fermions in the fundamental representation, with simulations carried out with a small lattice cutoff fixed in physical units, is enhanced when $N_f$  is increased from $2$ to $6$ -- by $40\%$ or more from inspection of the simulation results (Fig. \ref{ratio}). An enhancement of less than $40\%$ would require a significant downturn in $R_m$ (Fig. \ref{ratio}) before the eventual applicability of NLO $\chi$PT very likely bends it up again. Even the conservative lower bound of Eq. \ref{6fratio} indicates a substantial increase. This enhancement of $\pbp /F^3$ at $N_f = 6$ relative to $N_f = 2$, arising from distance scales between the confinement scale ($\sim 1/M_{\rho}$) and the lattice scale ($\sim (5M_{\rho})^{-1}$), appears even though the $N_f = 6$ theory is not yet walking. (The
running coupling at the lattice scale for $N_f = 6$ is still rather weak -- not much stronger than for $N_f = 2$.)

 For comparison, a perturbative $\MSbar$ computation of the enhancement of $\pbp /F^3$ at $N_f = 6$ relative to $N_f = 2$, based on the anomalous dimension of $\pbp$, gives an enhancement on the order of $1.05$ to $1.1$, depending on the order of perturbation theory \cite{LSD:2010}. This can be converted to the lattice scheme using the results of Ref. \cite{Aoki:2002iq} by multiplying by the ratio $Z^{\MSbar}_{6f} / Z^{\MSbar}_{2f} =  1.449(29) / 1.227(11) = 1.18(3)$. The result is a perturbative enhancement on the order of $20\% - 30\%$.

 It will be helpful to obtain results for smaller $m$ (with a larger lattice volume) and perhaps study the chiral extrapolations at NNLO. We are now exploring larger values of $N_f$ ($\rightarrow N_f^c$) and other gauge groups, and we will study the consequences of walking for quark and lepton mass generation and electroweak precision measurements.

We thank LLNL and the Multiprogrammatic and Institutional Computing
 program for time on the BlueGene/L supercomputer, and the Aspen Center
 for Physics.  This work was supported by the National Nuclear Security
 Agency and Office of Science (High Energy Physics), U.S. Department of
 Energy; and by the U.S. National Science Foundation.
 
 \vspace{-1mm}


\bibliography{lsd2f6f}

\end{document}